# A Universal Method to Transform Aromatic Hydrocarbon Molecules into Confined Carbyne inside Single-Walled Carbon Nanotubes

*Yingzhi Chen[1], Kunpeng Tang[1], Wendi Zhang[2], Huiju Cao[1], Hongwei Zhang[1], Yanghao Feng[1], Weili Cui[1], Yuan Hu[2,\*], Lei Shi[1,\*], Guowei Yang[1]*

[1] State Key Laboratory of Optoelectronic Materials and Technologies, Guangdong Basic Research Center of Excellence for Functional Molecular Engineering, Nanotechnology Research Center, School of Materials Science and Engineering, Sun Yat-sen University, Guangzhou, 510275, P. R. China

[2] School of Physical Science and Technology & Shanghai Key Laboratory of High-resolution Electron Microscopy, ShanghaiTech University, Shanghai 201210, P. R. China

**KEYWORDS:** confined carbyne, carbon chain, aromatic hydrocarbon molecule, single-walled carbon nanotubes, Raman Spectroscopy

**ABSTRACT:** Carbyne, a $sp^1$-hybridized allotrope of carbon, is a linear carbon chain with exceptional theoretically predicted properties that surpass those of $sp^2$-hybridized graphene and carbon nanotubes (CNTs). However, the existence of carbyne has been debated due to its instability caused by Peierls distortion, which limits its practical development. The only successful synthesis of carbyne has been achieved inside CNTs, resulting in a form known as confined carbyne (CC). However, CC can only be synthesized inside multi-walled CNTs, limiting its property-tuning capabilities to the inner tubes of the CNTs. Here, we present a universal method for synthesizing CC inside single-walled carbon nanotubes (SWCNTs) with diameter of 0.9-1.3 nm. Aromatic hydrocarbon molecules are filled inside SWCNTs and subsequently transformed into CC under low-temperature annealing. A variety of aromatic hydrocarbon molecules are confirmed as effective precursors for formation of CC, with Raman frequencies centered around 1861 cm$^{-1}$. Enriched (6,5) and (7,6) SWCNTs with diameters less than 0.8 nm are less effective than the SWCNTs with diameter of 0.9-1.3 nm for CC formation. Furthermore, resonance Raman spectroscopy reveals that optical band gap of the CC at 1861 cm$^{-1}$ is 2.353 eV, which is consistent with the result obtained using a linear relationship between the Raman signal and optical band gap. This newly developed approach provides a versatile route for synthesizing CC from various precursor molecules inside diverse templates, which is not limited to SWCNTs but could extend to any templates with appropriate size, including molecular sieves, zeolites, boron nitride nanotubes, and metal-organic frameworks.

INTRODUCTION

Carbyne is defined as an infinitely long linear carbon chain (LCC), or a sufficiently long LCC whose properties become independent of its length. The intrinsically linear structure of carbyne with $sp^1$-hybridizition has spurred extensive theoretical studies, which predict extraordinary properties[1-4], often surpassing those of $sp^2$-hybridized graphene and carbon nanotubes (CNTs). However, synthesizing carbyne presents a significant challenge, as it is inherently unstable due to Peierls distortion[5, 6]. Early efforts to explore carbyne focused on the synthesis of short LCCs, such as polyynes (characterized by alternating single and triple bonds)[7-9] and cumulenes (which feature successive double bonds)[10, 11]. Generally, cumulenes are even more unstable than polyynes[12]. Thus, synthesizing long polyynes has been a primary focus in the study of carbyne.

As the length of a polyyne increases, its instability becomes more pronounced. To address this, various stabilization strategies have been developed, including the attachment of chemical end groups[13-15], the formation of cyclic structures[16-18], and protection by CNTs[19-21]. Stabilizing polyynes by introducing a range of end groups—such as hydrogen[22-24], alkyl[13, 25, 26], aryl[14], trialkylsilyl[15], and even organometallic[27-29] groups—has enabled the synthesis of longer polyynes. Through these methods, polyynes consisting of up to 44[30], 48[15], 52[31], and 68[32] carbon atoms have been produced over time. However, these polyynes still exhibit length-dependent properties, meaning they do not qualify as true carbyne. Recently, on-surface synthesis has been employed to create metalated polyynes, which are derived from hexabromobenzene on an Ag(111) substrate via heat treatment[33]. Notably, a linear carbon chain containing 120 carbon atoms was synthesized on the surface of Au(111) through the demetallization of organometallic polyynes[34]. This structure is stable only in a vacuum, complicating its property investigation. In another advancement, cyclo[n]carbons were synthesized from bromocyclocarbon precursors by applying heating and/or voltage pulses via a scanning tunneling microscope tip. Cyclo[n]carbons with n = 10–26 have been successfully obtained using this method[16-18]. However, these structures, consisting of ≤ 26 carbon atoms, are polyynes rather than carbyne. Similarly, the formation of short LCCs has been observed through electron beam irradiation on graphene, which created LCCs between the holes in the graphene lattice[8, 35]. Additionally, LCCs have been observed when a CNT was stretched until it breaks into two pieces[36]. However, the length of LCCs is generally limited to a few nanometers—typically around 20 carbon atoms—and their lifetimes are short, usually lasting only tens of seconds. Consequently, it is crucial to develop methods for synthesizing long, stable LCCs that can be used for property measurements and potential applications under ambient conditions.

CNTs, characterized by their hollow structure and small diameter, serve as ideal templates for the growth of various one-dimensional nanomaterials, such as metal chains[37-39] and graphene nanoribbons[40-43]. Notably, the newly formed one-dimensional nanomaterials benefit from the protective environment provided by the CNTs, exhibiting greater stability compared to their free counterparts. Consequently, the confined synthesis of long LCCs with high stability using CNTs is highly anticipated. LCCs were first synthesized simultaneously with multi-walled CNTs through the arc-discharge method, resulting in LCCs exceeding 20 nm in length, as observed via high-resolution transmission electron microscopy (HRTEM)[19]. This indicates that the LCCs comprise more than 160 carbon atoms, significantly longer than the chemically terminated polyynes synthesized in solution or on substrate. Subsequently, a high-temperature treatment (~1500 °C) of CNTs was developed to produce long LCCs containing over 6000 carbon atoms, achieved using double-walled carbon nanotubes (DWCNTs) with an average inner diameter of 0.8 nm[21]. It was determined that these LCCs are sufficiently long for their properties to become independent of length, marking the first experimental confirmation of carbyne. Since this long LCC exists exclusively inside the CNT, it is referred to as confined carbyne (CC) to differentiate it from free carbyne, which has yet to be confirmed. Interestingly, the properties of CC are not length-dependent but are influenced by the inner tubes of the DWCNTs through interactions between the CC and the nanotube[44, 45]. This is particularly true for CCs inside DWCNTs with inner diameters ranging from 0.6 to 0.8 nm. To mitigate the influence of the nanotube on the CC, single-walled CNTs (SWCNTs) with larger diameters (>1.1 nm) were employed for CC synthesis via high-temperature treatment[46, 47]. Unfortunately, while CC was successfully produced, the SWCNTs were found to have transformed into DWCNTs to counteract the instability of SWCNTs at high temperatures. Thus, there is a pressing need to develop methods for growing CC inside SWCNTs.

Previously, short polyynes were introduced into SWCNTs in a solvent and subsequently annealed at temperatures above 800 °C to coalesce neighboring short polyynes into a long polyyne[48, 49]. In contrast, CC@SWCNT can be extracted from CC@DWCNT by applying tip-ultrasonication, followed by separation from the outer tubes and unextracted CC@DWCNTs using density gradient ultracentrifugation[50]. Although both methods can yield CC@SWCNT, the overall yield of CC remains low. More critically, the CCs remain confined inside thin SWCNTs, significantly affecting their properties. Therefore, there is an urgent need to develop a new route for the efficient synthesis of CC inside SWCNTs, particularly those with diameters greater than 0.8 nm. During the preparation of the manuscript, surfactants were reported as precursors to synthesize CC inside SWCNTs with diameter of >0.95 nm, resulting in weak interaction between the CC and SWCNTs[51].

Here, we propose a novel and universal approach, termed the confined transformation method, for preparing CC inside SWCNTs. This method involves transforming aromatic hydrocarbon molecules into CC by annealing them inside SWCNTs with diameter of 0.9-1.3 nm at relatively low temperatures (<600 °C). These SWCNTs facilitate the easy insertion of aromatic hydrocarbon precursors, and the one-dimensional confined space promotes a linear transformation of the molecules into CC. Resonance Raman spectroscopic studies indicate that the optical band gap of the CC is approximately 2.353 eV, which aligns well with the linear relationship between the optical band gap and the Raman frequency of CC established in our previous research. HRTEM observations confirmed the existence of the CC inside the SWCNT. In our experiments, we successfully transformed nine different aromatic hydrocarbon molecules into CC, suggesting that this method is universal and can accommodate a variety of precursor molecules, not limited to aromatic hydrocarbons.

## RESULTS AND DISCUSSION

As illustrated in Figure 1, we propose a universal method for producing CC inside SWCNTs. In this method, sublimated precursor molecules are first inserted into the SWCNTs, followed by a subsequent annealing process that facilitates the transformation of these molecules into CC, leveraging the confined space. To investigate the transformation mechanism from aromatic hydrocarbon molecules to CC, we utilized a total of nine different precursor molecules. It is important to note that both the sublimation temperature and the optimal annealing temperature varied depending on the specific type of molecule used.

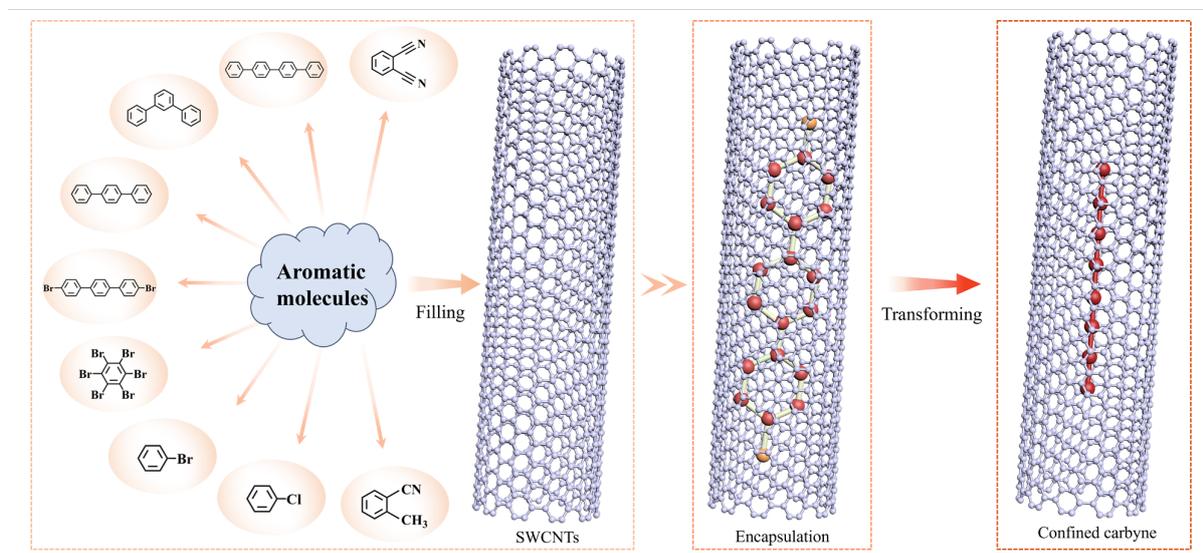

**Figure 1.** Schematic of confined transformation method. Various aromatic hydrocarbon molecules are chosen to be filled inside SWCNT and then transformed into CC under annealing.

Taking 4,4"-dibromo-p-terphenyl (DBTP) as an example, the molecules were filled and transformed inside SWCNTs with diameters ranging from 0.7 to 1.3 nm. Due to the size of DBTP, the molecules could only be encapsulated in the SWCNTs with diameter greater than 0.9 nm. The DBTP molecules were sublimated at 200 °C and subsequently annealed at temperatures ranging from 400 to 700 °C for 60 minutes, or at 500 °C for durations between 10 and 60 minutes. As shown in Figures 2a and 2b, the Raman signals at approximately 269 cm$^{-1}$ is attributed to the radial breathing mode (RBM) of the SWCNTs. According to the equation D=234/($\omega_{RBM}$-10)[52], where D is the SWCNT diameter in nm and $\omega_{RBM}$ is the RBM frequency in cm⁻¹, the calculated diameter of the observed SWCNTs is around 0.9 nm, which is suitable for molecule encapsulation. A new Raman signal appearing at approximately 1858 cm⁻¹ after annealing is assigned to the CC-band[53, 54], which corresponds to the formation of CC, i.e., the transformation from molecule to CC. This CC-band intensity peaks at 500 °C, much lower than the 1460 °C typically used in the high-temperature treatment method[21] and also lower than the 800 °C used for coalescing polyyne molecules into longer chains inside SWCNTs[48]. Interestingly, the CC-band frequency is at a relatively higher frequency (1858 cm⁻¹) compared to previous studies where it was observed between 1790-1855 cm⁻¹ for CC inside DWCNTs[21, 48, 55]. This shift can be attributed to the weaker interaction between CC and the larger SWCNTs compared to the smaller inner tubes of DWCNTs. Moreover, the length of the CC in this case could be shorter than those observed previously, although this requires further investigation.

To investigate the role of diameter in the formation of CC, (6,5) and (7,6) enriched CoMoCAT SWCNTs were used as templates. While these chirality-enriched samples contain a great fraction of small SWCNTs with diameters less than 0.8 nm, they also have a smaller fraction of larger nanotubes, and their average diameters are smaller than those in the standard CoMoCAT sample. Although CC was able to be formed inside these chirality-enriched SWCNT samples (Figures 2c and S1), the significantly lower intensity of the CC band suggests that CC is more likely synthesized inside the larger-diameter SWCNTs. This is because the (6,5) and (7,6) SWCNTs, with diameters less than 0.8 nm, are too narrow to encapsulate the DBTP precursor molecules effectively. In contrast, no CC formation was observed when SWCNTs with diameters of around 1.3 nm were used (Figure S1). Therefore, SWCNTs with diameters between 0.9 and 1.3 nm are the optimal size range for encapsulating precursor molecules and facilitating their transformation into CC.

A Raman contour mapping was conducted to evaluate the sample annealed at 500 °C, as this sample exhibited a more heterogeneous distribution of the CC-mode compared to the one annealed at 600 °C. As shown in Figure 2d, the uneven distribution of Raman intensity for the CC-band indicates that the yield of CC is not uniform. This is attributed to the random distribution of SWCNTs with varying diameters and the changing filling ratio of molecules inside the SWCNTs. The statistics of the intensity ratio between the CC-mode and the $G^+$-mode ($I_{CC}/I_{G^+}$) and the Raman frequency distribution of the CC-mode are summarized in Figures 2e and 2f, respectively. In general, the yield of CC is still lower than that achieved through high-temperature treatments in previous studies[21, 47, 55]. However, CC is consistently observed throughout the sample, demonstrating the effectiveness of the method. Notably, the frequency of the CC-mode varies from 1850 to 1867 cm⁻¹ when sampling at different positions, likely due to the diameter variation of the SWCNTs.

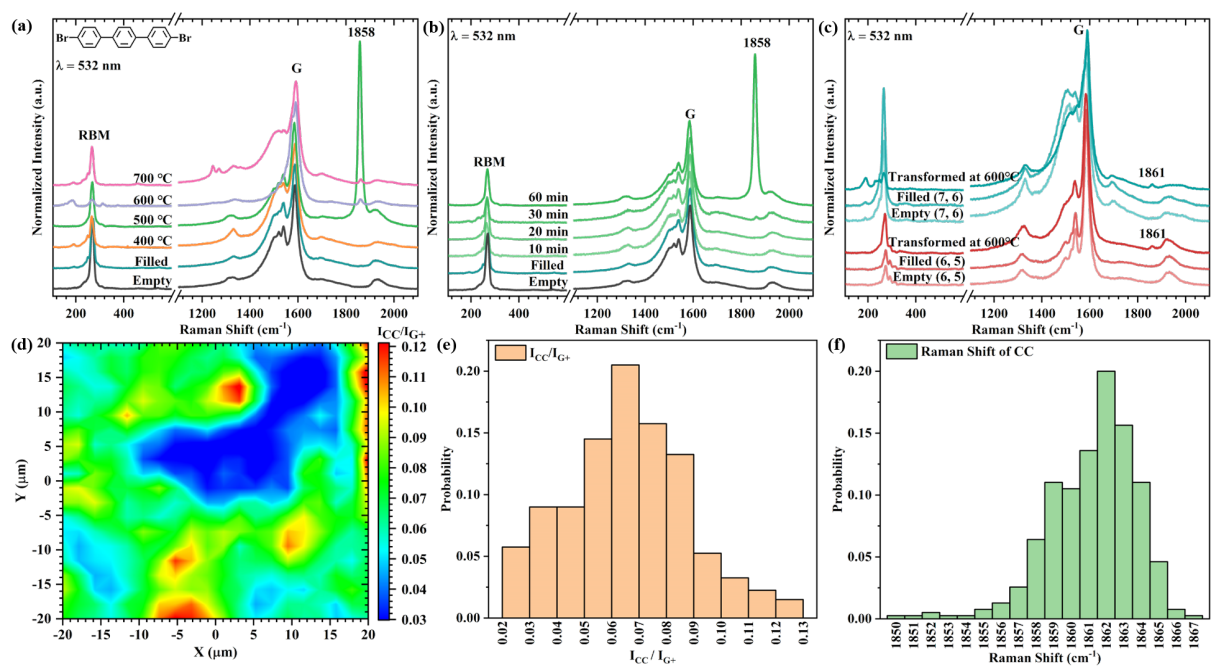

**Figure 2.** Transformation of DBTP into CC inside SWCNTs. Raman spectra of empty, filled and transformed samples at different temperatures (a) for different durations at 500 °C (b). (c) Raman spectra of empty, filled and transformed samples using (6,5) and (7,6) enriched SWCNTs as templates. (d) Raman contour mapping and (e) Statistical histogram of normalized CC-mode to the G-band of CC@SWCNT transformed from DBTP at 500 °C. (f) Statistical histogram of Raman frequency in (d).

To obtain comprehensive information about the sample annealed at 500 °C, several discrete lasers were employed for Raman spectroscopic studies. Note that switching lasers did not alter the excitation position on the sample, ensuring that the testing area remained fixed at approximately 1.5 micrometers in diameter. As shown in Figure 3a, the highest intensity of the CC-mode was observed in Raman spectra excited by a 532 nm laser, indicating that the laser energy is close to the optical band gap of the CC. A slightly shift in the CC-mode is observed when excited by a 561 nm laser, attributed to the variation in the optical band gap with CC of different lengths and inside different SWCNTs. To determine the exact optical band gap, a tunable laser was utilized for excitation. As illustrated in Figure 3b, the intensity of the CC-mode at 1861 cm⁻¹ significantly changes with the excitation laser. The resonance Raman excitation profile in Figure 3c shows that the $I_{CC}/I_{G^+}$ ratio increases from 0.26 to 78.29 within a laser energy range of 2.2 to 2.4 eV. The enhancement factor reaches as high as 310 due to the resonance effect. The highest intensity of the CC-mode is excited by a 527 nm laser, indicating that the optical band gap of the CC at 1861 cm⁻¹ is 2.353 eV. Previously, a linear relationship between the optical band gap and Raman frequency was established[30-32, 56-58]. Our results were plotted alongside this linear function, as shown in the inset of Figure 4c. The experimentally obtained value of 2.353 eV in the solid cycle is highly consistent with the predicted value of 2.357 eV in the open cycle, further validating the effectiveness of this relationship and allowing its application for further analysis. The Raman frequency range from 1855 to 1865 cm⁻¹, when applied to the fitting linear line, allows for the prediction of the optical band gap, which ranges from 2.3 to 2.4 eV. This range falls well within the gap (marked by the dashed rectangle in Figure 4c) between the CC inside DWCNT and the polyynes in the solvent.

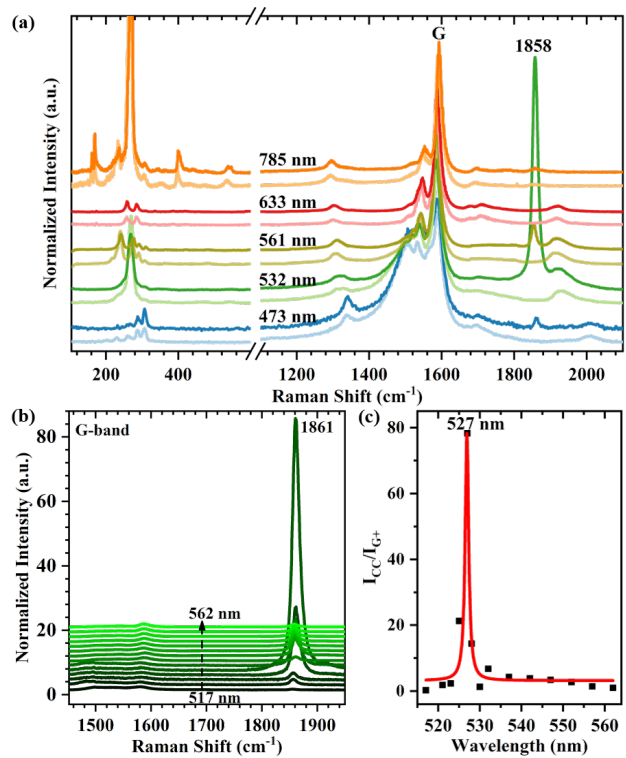

Figure 3. (a) Raman spectra of empty SWCNTs (light line) and CC@SWCNT (dark line) transformed from DBTP at 500 °C excited by lasers with different wavelengths. (b) Resonance Raman spectra excited by a tunable laser with wavelengths from 517 to 562 nm. (c) Resonance Raman excitation profile of the CC-mode at 1861 cm$^{-1}$.

To elucidate the transformation mechanism from aromatic hydrocarbon molecules to CC, we compared eight additional molecules alongside DBTP. A detailed study was conducted on the temperature and duration-dependent synthesis for each molecule, with the corresponding Raman spectra presented in Figures S2-S3 of the supporting information. The Raman spectra of the optimized samples are summarized in Figure 4. Note that the Raman frequency of the CC-band not only varies with different precursor molecules and their respective annealing temperatures and durations but is also influenced by the uniformity of the sample and the excitation laser energy. In comparison to DBTP, para-terphenyl was converted to CC with a lower yield, indicating that bromination plays a significant role in the opening of the benzene ring, as dehydrogenation is more challenging than debromination. The yield of CC further decreases for meta-terphenyl due to its slightly wider structure, necessitating larger SWCNTs for accommodation, which are present in smaller fractions in the sample. Similarly, the filling efficiency of para-quaterphenyl is lower than that of para-terphenyl, resulting in a reduced yield of CC. While bromobenzene and chlorobenzene could be theoretically filled inside SWCNTs more easily, the results show a less intense CC band. In contrast, higher CC bands are observed for 1,2-dimethylbenzene and o-tolunitrile, which feature neighboring functionalizations. This suggests that although a single functionalization on the benzene ring facilitates ring opening, further dehydrogenation can be enhanced through the introduction of neighboring functional groups. Additionally, hexabromobenzene exhibits the lowest transformation yield due to its lack of hydrogen, which is crucial for stabilizing the formed CC by terminating with hydrogen at the ends. Based on this analysis, we propose a fundamental guideline for identifying optimal precursor molecules for transformation into CC: a molecule should contain hydrogen and neighboring functionalizations, although it does not necessarily need to have a benzene ring structure. Following this guideline, we anticipate the discovery of better precursor molecules in the near future.

Previously, a linear relationship was established between the Raman frequency and the inverse number of carbon atoms in a short linear carbon chain, with the Raman frequency saturating for longer chains[21, 32, 57, 59-63], as illustrated in Figure 4b. When the CC is inside a SWCNT, Raman frequency of the CC is not only dependent on the length, but also varies with the diameter of the hosted SWCNT due to their interactions. In our study, Raman frequency of the CC is observed to range from 1857-1867 cm$^{-1}$, as indicated by the two short dashed horizontal lines in Figure 4b. On one hand, if the CC behaves more like a free chain—meaning there is minimal interaction with the SWCNTs—it would consist of more than 100 carbon atoms, corresponding to a length greater than 10 nm. On the other hand, if the interaction significantly influences the modulation of the Raman frequency of the CC, it would follow a linear relationship at the lower end, resulting in a CC with approximately 40 carbon atoms, or about 4 nm in length. To resolve this ambiguity, direct observation using HRTEM is necessary.

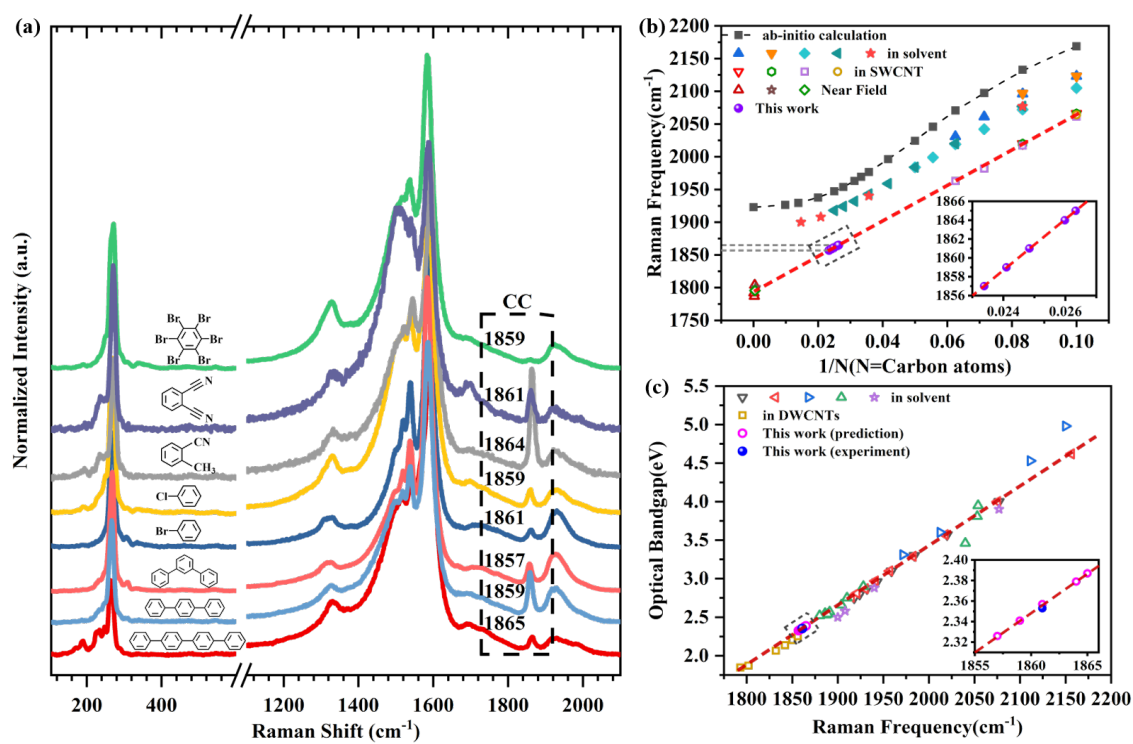

**Figure 4.** (a) Raman spectra of CC@SWCNT transformed from different precursor molecules excited by a laser with wavelength of 532 nm. (b) Raman frequency of the samples in (a) as a function of inverse length, given by the number *N* of carbon atoms. Parts of the data are taken from references of *ab initio* calculations[21] and for samples in solvent[32, 57, 59, 60], inside SWCNTs[61-63], and inside DWCNTs via near-field measurements[21]. Inset: the enlarged part of the data obtained in this work. (c) Optical band gap as a function of Raman frequency of samples in (a). The data in open cycles are obtained from the calculation using equation $E=0.0076\omega_{CC}-11.81$, and the data in solid cycle is taken from resonance Raman spectroscopic study. The rest data are taken from references for samples in solvent[30-32, 56, 57] and inside DWCNTs[58]. Inset: the enlarged part of the data in this work.

The successful formation and morphology of the CC confined inside SWCNTs were directly observed using HRTEM. As shown in Figures 5a-5c, a CC is confined inside a SWCNT with a diameter of 0.90 nm. Unlike the fully linear CC structure observed inside DWCNTs with an inner diameter of approximately 0.7 nm in our previous study[21], the CC inside the SWCNT is not a complete straight linear structure. One end of the CC is free to move, while the other end is coiled into a circular shape. This structure, as seen in the HRTEM image, could be considered a short graphene nanoribbon. However, graphene nanoribbons inside SWCNTs typically adopt a twisted structure, which rotates dynamically under electron irradiation. In our case, no twisting was observed over time, which rules out the possibility of a graphene nanoribbon structure. Additionally, another CC was observed inside a wider SWCNT with a diameter of 1.06 nm (Figures 5e-5g). This CC exhibited two free linear structures at both ends and a looped shape in the middle. The linear structure on the left was initially positioned near the upper sidewall of the SWCNT, but after a few seconds, it moved towards the lower sidewall. The looped shape is attributed to the sufficient space within the SWCNT, which allows the CC to move not only along the tube but also radially within it. The lengths of these two CCs range from approximately 1.4 to 2.5 nm. If there is no interaction between the CC and the large SWCNT, the Raman frequency of the CC should adhere to the upper linear fitting line depicted in Figure 4b, resulting in frequencies exceeding 1900 cm$^{-1}$. However, this is not the case. Therefore, our results align more closely with the interaction model rather than the free-chain model. Thus, the interaction between the CC and the SWCNT remains evident even in the case of a larger diameter SWCNT, as the CC is not centered in the nanotube but is instead typically located closer to one of the walls, as seen in the HRTEM observations. While this interaction is less pronounced than in narrower tubes, it still influences the properties of the CC. Further studies on the chain-length-dependent and SWCNT-diameter-dependent properties of individual CC@SWCNTs should be conducted using tip-enhanced Raman spectroscopy to assess the respective contributions of the CC and the SWCNT in modulating the properties of the CC.

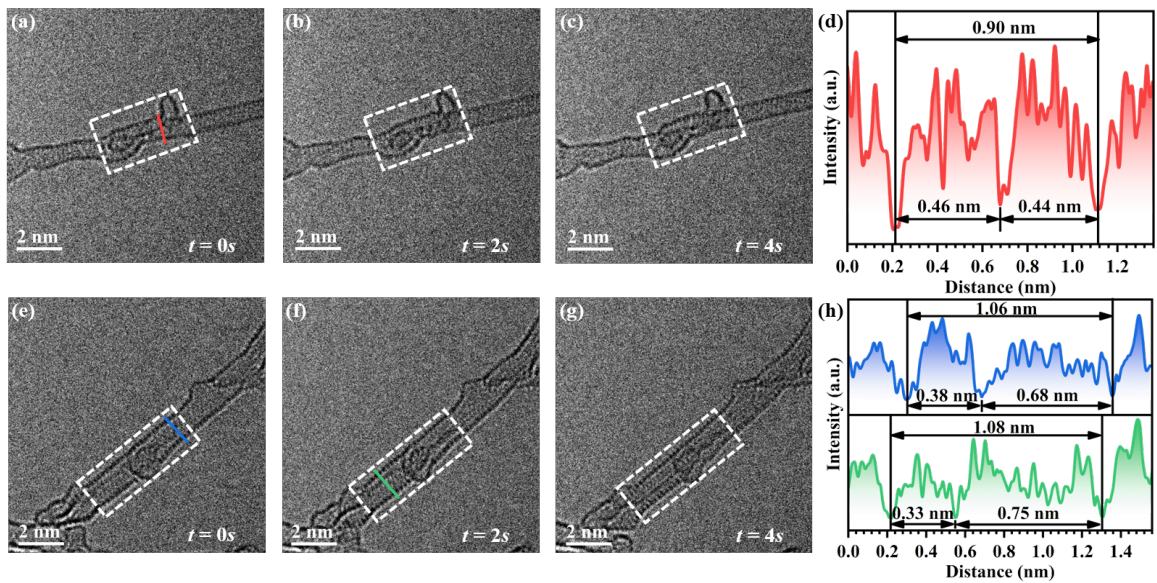

**Figure 5.** (a-c) HRTEM images of a CC@SWCNT taken at different times. (d) Contrast profile at the cross-section marked in (a). (e-g) HRTEM images of another CC@SWCNT taken at different times. (d) Contrast profiles at the cross-sections marked in (e) and (f).

## CONCLUSIONS

In this study, we have demonstrated a novel and universal method for synthesizing CC inside SWCNTs, expanding the possibilities for carbyne synthesis beyond the limitations of multi-walled CNTs. By filling aromatic hydrocarbon molecules inside SWCNTs and subjecting them to low-temperature annealing, we successfully converted a variety of aromatic precursors into CC, with Raman frequencies of 1857 - 1867 $cm^{-1}$. Our results indicate that SWCNTs with diameter greater than 0.9 nm are able to synthesize CC. The optical properties of CC were also investigated, with resonance Raman spectroscopy revealing an optical band gap of 2.353 eV at 1861 $cm^{-1}$, supporting our previously established a linear relationship between Raman frequency and optical band gap. Furthermore, HRTEM analysis showed that the CC inside large SWCNTs exhibited unique morphologies: the CC displayed both linear and looped forms, with the linear segment located near the sidewalls of the SWCNT. This suggests that the interaction between the CC and the nanotube still plays a role in modulating the properties of the CC, even in larger-diameter SWCNTs. Our approach is not restricted to SWCNTs and can be extended to a wide range of templates, including molecular sieves, zeolites, boron nitride nanotubes, and metal-organic frameworks, providing a versatile and scalable route for the synthesis of CC from various precursor molecules. This work not only provides a pathway for synthesizing CC in a more accessible and versatile manner but also advances the understanding of its structure and properties when confined inside large SWCNTs.

## METHODS

Standard CoMoCAT SWCNTs, (6,5) and (7,6) enriched CoMoCAT SWCNTs (Sigma-Aldrich, manufactured by SouthWest NanoTechnologies Inc.) were used as template for synthesis of CC. The SWCNTs were used as obtained without further purification. The SWCNTs were opened by thermal treatment in air at 400°C for 1 hour. The opened SWCNTs together with precursor molecules, such as DBTP, were sealed together in an ampoule in a dynamic vacuum (pressure < $10^{-3}$ Pa). The sealed ampoules were heated at temperatures of 200-400°C for 3 days to fill the precursor molecules into the SWCNTs. The sublimation temperature varies with precursor molecules. The filled samples were then taken out from the ampoules and subsequently annealed at temperatures ranging from 400 to 700°C for durations from 10-60 min in a dynamic vacuum environment (a pressure < $10^{-3}$ Pa). Such annealing process facilitated the transformation of the precursor molecules into CC inside the SWCNTs.

The samples underwent characterization using Raman spectroscopy (TriVista 557, Princeton Instruments, equipped with a liquid nitrogen-cooled CCD detector, a 50× objective, and a 600 gr/mm grating) with a laser wavelength of 473, 532, 561, 633, and 785 nm to excite the CC. To minimize the heating effect, the laser power was maintained below 1 mW. For the resonance Raman spectroscopy (TriVista 777, Princeton Instruments, equipped with a liquid nitrogen-cooled CCD detector, a 50× objective, three 1800 gr/mm grating) a tunable laser (C-WAVE GTR) was applied for excitation. All spectra were calibrated against the Rayleigh scattering line and normalized to the intensity of the G-band of SWCNTs. To directly observed the formation of CC, the sample was characterized using aberration-corrected HRTEM (JEOL Grand ARM300F) at an operating voltage of 80 kV. Each image was taken with an exposure time of 1 second.

**Supporting Information**

S1. Transformation of DBTP into CC inside different types of SWCNTs.

S2&S3. Transformation of different molecules into CC inside CoMoCAT SWCNTs.

The following file are available free of charge.

[Supplementary Information](#)

AUTHOR INFORMATION

**Corresponding Author**

**Yuan Hu** – School of Physical Science and Technology & Shanghai Key Laboratory of High-resolution Electron Microscopy, ShanghaiTech University, Shanghai 201210, P. R. China. E-mail: *huyuan@shanghaitech.edu.cn

**Lei Shi** – State Key Laboratory of Optoelectronic Materials and Technologies, Guangdong Basic Research Center of Excellence for Functional Molecular Engineering, Nanotechnology Research Center, School of Materials Science and Engineering, Sun Yat-sen University, Guangzhou, 510275, P. R. China. E-mail: *shilei26@mail.sysu.edu.cn

Notes

The authors declare no competing financial interest.

ACKNOWLEDGMENTS

This work supported by National Natural Science Foundation of China (52472059, 22401298, 22302123) and State Key Laboratory of Optoelectronic Materials and Technologies (OEMT-2022-ZRC-01).

# Supporting Information

# A Universal Method to Transform Aromatic Hydrocarbon Molecules into Confined Carbyne inside Single-Walled Carbon Nanotubes

*Yingzhi Chen[1], Kunpeng Tang[1], Wendi Zhang[2], Huiju Cao[1], Hongwei Zhang[1], Yanghao Feng[1], Weili Cui[1], Yuan Hu[2,\*], Lei Shi[1,\*], Guowei Yang[1]*

[1] State Key Laboratory of Optoelectronic Materials and Technologies, Guangdong Basic Research Center of Excellence for Functional Molecular Engineering, Nanotechnology Research Center, School of Materials Science and Engineering, Sun Yat-sen University, Guangzhou, 510275, P. R. China

[2] School of Physical Science and Technology & Shanghai Key Laboratory of High-resolution Electron Microscopy, ShanghaiTech University, Shanghai 201210, P. R. China.

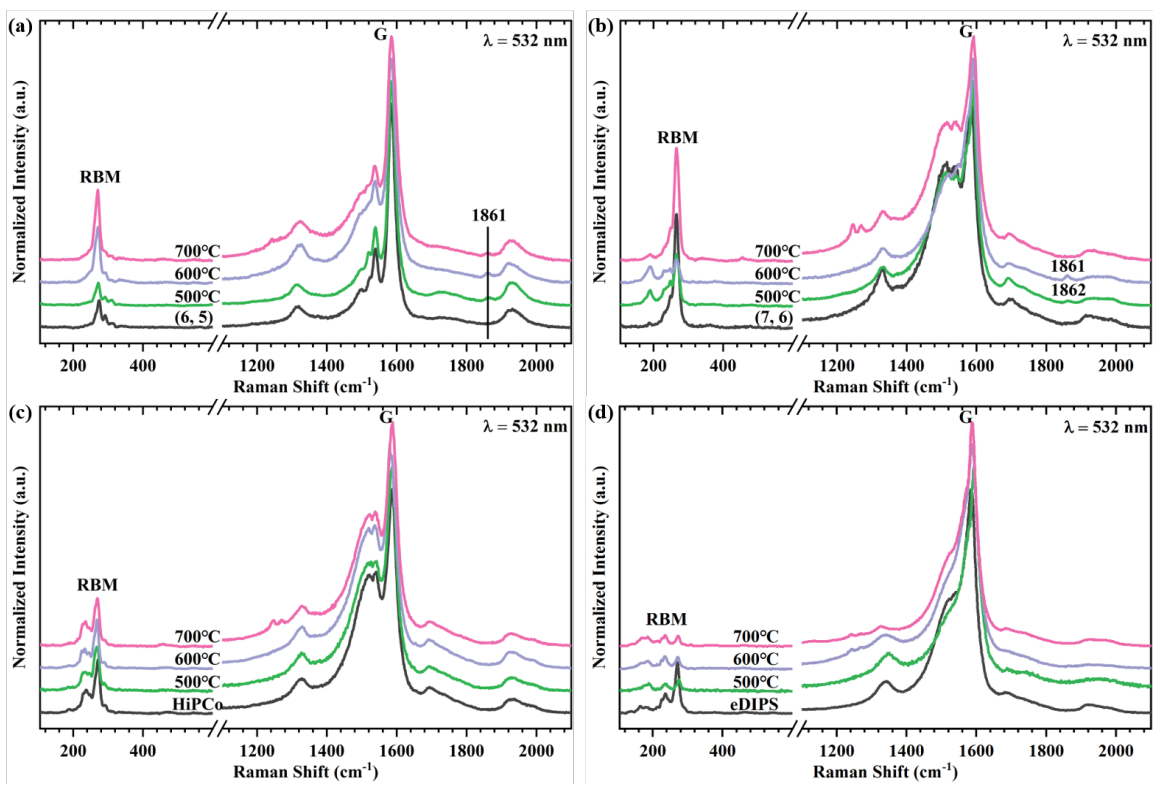

**Figure S1.** Raman spectra of empty and temperature-dependent DBTP-transformed CCs inside different types of SWCNTs. (a) (6,5) enriched SWCNT, (b) (7,6) enriched SWCNT, (c) HiPCo SWCNT, (d) eDIPS-1.3 nm SWCNT.

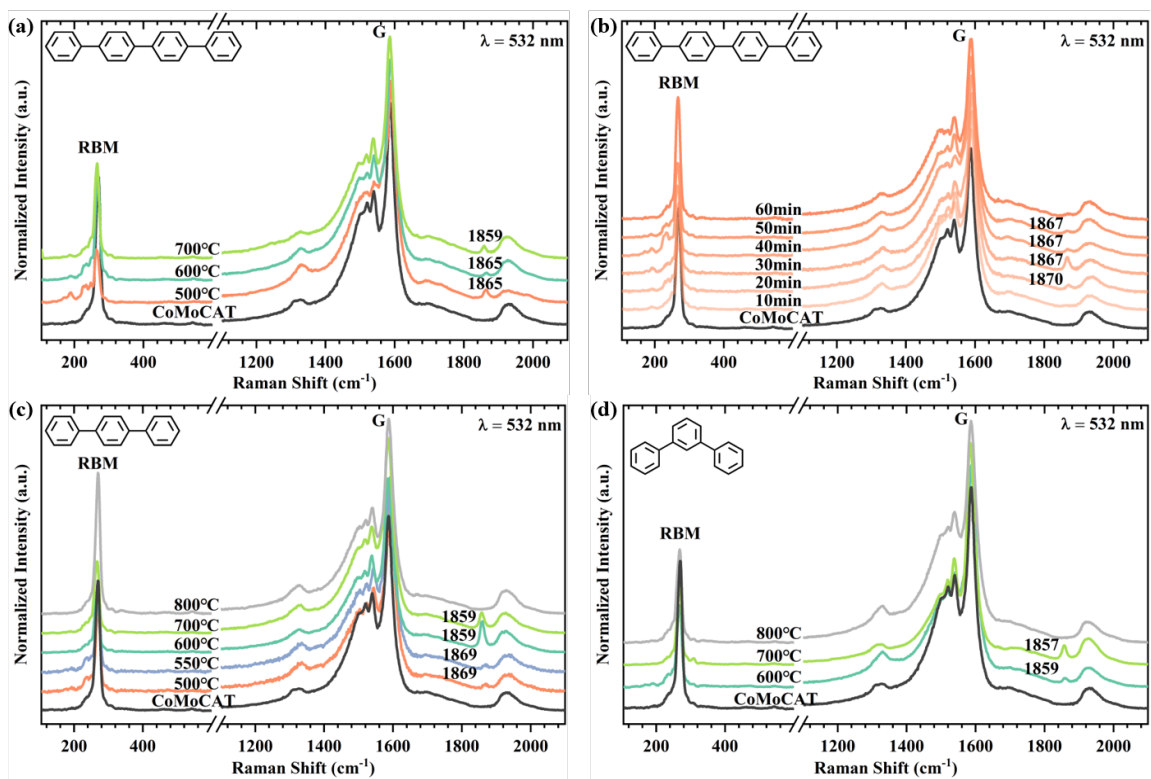

**Figure S2.** Raman spectra of empty CoMoCAT SWCNTs and para-quaterphenyl transformed CC inside CoMoCAT SWCNTs (a) at different temperature and (b) for different durations at 500 °C. Raman spectra of empty CoMoCAT SWCNTs and (c) para-terphenyl and (d) meta-terphenyl transformed CC inside CoMoCAT SWCNTs at different temperature.

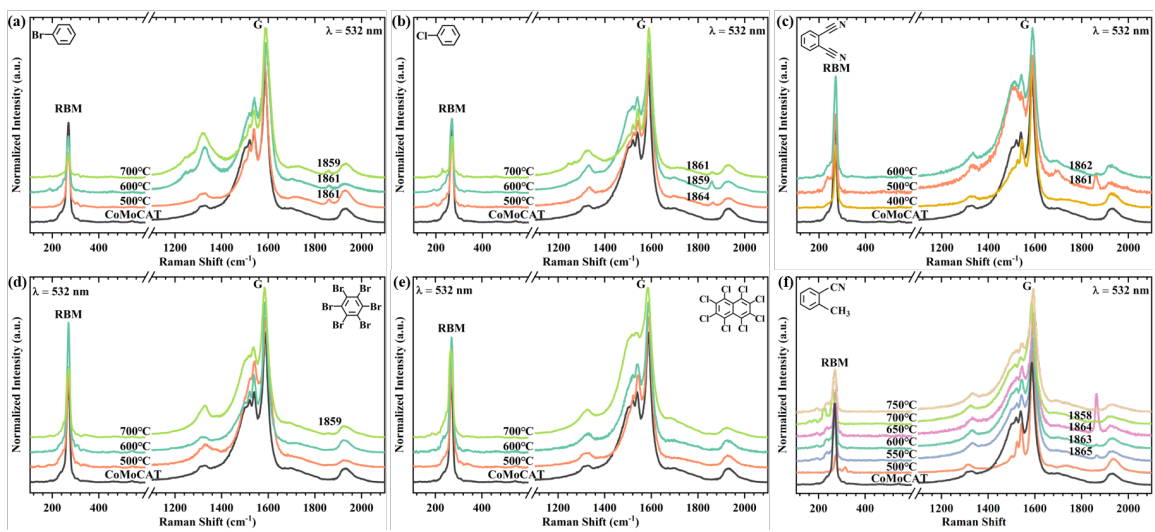

**Figure S3.** Raman spectra of empty CoMoCAT SWCNTs and different precursor molecules transformed CC inside CoMoCAT SWCNTs. (a) Bromobenzene, (b) Chlorobenzene, (c) 1,2-Dimethylbenzene, (d) Hexabromobenzene, (e) Octachloronaphthalene, (f) O-tolunitrile.